\documentclass[aps,prb,twocolumn,showpacs,amsmath,amssymb]{revtex4}
\usepackage{graphicx}
\usepackage{bm}

\def\br{ \bm{r} }
\def\bk{ \bm{k} }
\def\bp{ \bm{p} }
\def\bq{ \bm{q} }
\def\bB{ \bm{B} }

\def\bM{ \bm{M} }
\def\bgam{ \bm{\gamma} }

\def\im{ \,\mathrm{Im}\, }
\def\re{ \,\mathrm{Re}\, }

\def\tr{\,\mathrm{tr}}

\begin{document}

\title{On the spin susceptibility of noncentrosymmetric superconductors}

\author{K. V. Samokhin}

\affiliation{Department of Physics, Brock University,
St.Catharines, Ontario L2S 3A1, Canada}
\date{\today}

\begin{abstract}
We calculate the spin susceptibility of a superconductor without
inversion symmetry, both in the clean and disordered cases. The
susceptibility has a large residual value at zero temperature,
which is further enhanced in the presence of scalar impurities.
\end{abstract}


\maketitle

\section{Introduction}
\label{sec: Intro}

The discovery of superconductivity in CePt$_3$Si (Ref.
\onlinecite{Bauer04}) has renewed interest, both experimental and
theoretical, in the properties of superconductors without
inversion symmetry. The list of such superconductors has been
steadily growing and now also includes UIr (Ref.
\onlinecite{Akazawa04}), CeRhSi$_3$ (Ref. \onlinecite{Kimura05}),
CeIrSi$_3$ (Ref. \onlinecite{Sugitani06}), Y$_2$C$_3$ (Ref.
\onlinecite{Amano04}), Li$_2$(Pd$_{1-x}$,Pt$_x$)$_3$B (Ref.
\onlinecite{LiPt-PdB}), and other materials.

A distinctive feature of noncentrosymmetric crystals is that the
spin-orbit (SO) coupling qualitatively changes the nature of
single-electron states, namely it lifts spin degeneracy and splits
energy bands almost everywhere in the Brillouin zone. This has
important consequences for superconductivity. In the limit of
strong SO coupling, the Cooper pairing between the electrons with
opposite momenta occurs only if they are from the same
nondegenerate band. According to Ref. \onlinecite{SZB04}, this is
what happens in CePt$_3$Si, where the SO band splitting ranges
from 500K to 2000K and thus exceeds the critical temperature
$T_c=0.75$K by orders of magnitude. The same is likely to be the
case in other materials, for instance Li$_2$Pd$_3$B and
Li$_2$Pt$_3$B, see Ref. \onlinecite{LP05}.

The pairing interaction in the strong SO coupling case is most
naturally introduced using the exact band
states,\cite{SZB04,GR01,SC04,Min04} which take into account all
the effects of the crystal lattice potential and the SO coupling.
In the band representation, the superconducting order parameter is
represented by a set of complex functions, one for each band,
which are coupled, e.g. due to the interband tunneling of the
Cooper pairs or the impurity scattering.\cite{MS07} Since the
bands are nondegenerate, the pairing symmetry is peculiar: While
the Pauli principle dictates that each order parameter is an odd
function of momentum, the gap symmetry, in particular the location
of the gap nodes, is determined by one of the even representations
of the crystal point group.

If the band splitting is smaller than the superconducting critical
temperature, then the effects of the SO coupling can be treated
perturbatively, in particular the pairing Hamiltonian can be
constructed using the basis of the pure spinor states unaffected
by the SO coupling. This approach to the theory of
noncentrosymmetric superconductivity was introduced in Ref.
\onlinecite{Edel89} and developed further in Refs.
\onlinecite{FAKS04,FAMS05}.

One of the most peculiar properties of noncentrosymmetric
superconductors is a strongly anisotropic spin susceptibility with
a large residual component at zero
temperature.\cite{Bul76,Edel89,GR01,Yip02,FAKS04,FAS04,Sam05,Fuji06}
The latter comes from the field-induced virtual interband
transitions and is almost unchanged across $T_c$. Its magnitude
depends on the electron structure and is considerably smaller than
the normal-state susceptibility, see Sec. \ref{sec: chi clean}
below. On the other hand, the experiments in CePt$_3$Si (Ref.
\onlinecite{Yogi06}) have detected essentially no change in the
Knight shift below $T_c$, for all directions of the applied field.
One possible explanation is that the relative magnitude of the
interband contribution increases due to the combined effect of
interactions and peculiarities of the electron
structure.\cite{Fuji06}

The goal of this article is to explore an alternative mechanism
whereby the residual spin susceptibility is enhanced by disorder.
It is well known that the SO impurities play an important role in
usual, i.e. centrosymmetric singlet, superconductors, in which
experiments also revealed a large residual Knight shift, see, e.g.
Ref. \onlinecite{HK71} and the references therein. This is
incompatible with the predictions of the Bardeen-Cooper-Schrieffer
(BCS) theory, even when the SO coupling with the crystal lattice
potential is taken into account,\cite{Gor65} and can be explained
by the presence of spin-reversing scattering at the sample
boundaries or impurities.\cite{Ferrell59,Anders59,AG62} We
calculate the spin susceptibility of a noncentrosymmetric
superconductor in the presence of scalar disorder. In contrast to
Refs. \onlinecite{Edel89} and \onlinecite{Bul76}, we focus on the
strong SO coupling limit and employ the band representation of the
pairing Hamiltonian.

The article is organized as follows: In Sec. \ref{sec: clean} we
calculate the spin susceptibility in the clean case. In Sec.
\ref{sec: impurities}, we introduce the impurity scattering in the
band representation and derive general expressions for the
susceptibility in the disordered case. In Sec. \ref{sec: residual
chi}, we calculate the residual susceptibility in some simple
models, in which the effects of impurities can be worked out
analytically. We have also included an Appendix with the
derivation of the effective SO Hamiltonian for band electrons.

\section{Clean case}
\label{sec: clean}

In a noncentrosymmetric crystal with SO coupling the electron
bands are nondegenerate. The formal reason is that without the
inversion operation one cannot, in general, have two orthogonal
degenerate Bloch states at the same wave vector $\bk$. In the
limit of zero SO coupling there is an additional symmetry in the
system -- the invariance with respect to arbitrary rotations in
spin space -- which preserves two-fold degeneracy of the bands.
Let us consider a single band of electrons with the dispersion
given by $\epsilon(\bk)$, and turn on the SO coupling with the
crystal lattice. The system Hamiltonian can be written in the
following form:
\begin{equation}
\label{H SO eff}
    H_0=\sum\limits_{\bk,\alpha\beta}[\epsilon(\bk)\delta_{\alpha\beta}+
    \bgam(\bk)\bm{\sigma}_{\alpha\beta}]
    a^\dagger_{\bk\alpha}a_{\bk\beta}.
\end{equation}
Here $\alpha,\beta=\uparrow,\downarrow$ label the spin states,
$\bm{\sigma}$ are the Pauli matrices, and the sum over $\bk$ is
restricted to the first Brillouin zone. The ``bare'' band
dispersion satisfies $\epsilon(-\bk)=\epsilon(\bk)$,
$\epsilon(g^{-1}\bk)=\epsilon(\bk)$, where $g$ is any operation
from the point group $\mathbb{G}$ of the crystal. The SO coupling
is described by the pseudovector function $\bgam(\bk)$, which has
the following symmetry properties: $\bgam(\bk)=-\bgam(-\bk)$,
$g\bgam(g^{-1}\bk)=\bgam(\bk)$. The derivation of Eq. (\ref{H SO
eff}) is outlined in the Appendix.

For the tetragonal group $\mathbb{G}=\mathbf{C}_{4v}$, which
describes the point symmetry of CePt$_3$Si, CeRhSi$_3$ and
CeIrSi$_3$, the SO coupling has the following form:
\begin{equation}
\label{gamma C4v}
    \bgam(\bk)=\gamma_\perp[\bm{\phi}_{E,u}(\bk)\times\hat z]
    +\gamma_\parallel\phi_{A_2,u}(\bk)\hat z.
\end{equation}
Here $\gamma_\perp$ and $\gamma_\parallel$ are constants, and
$\bm{\phi}_{E,u}$ and $\phi_{A_2,u}$ are the odd basis functions
of the irreducible representations $E$ (two-dimensional) and $A_2$
(one-dimensional), respectively. The representative polynomial
expressions for the basis functions are
$\bm{\phi}_{E,u}(\bk)\propto(k_x,k_y)$ and
$\phi_{A_2,u}(\bk)\propto k_xk_yk_z(k_x^2-k_y^2)$. A particular
two-dimensional case of Eq. (\ref{gamma C4v}) with
$\bgam(\bk)=\gamma_\perp(\bk\times\hat z)$ is known as the Rashba
model\cite{Rashba60} and has been extensively used to study the
effects of SO coupling in semiconductor heterostructures. For the
cubic group $\mathbb{G}=\mathbf{O}$, which describes the point
symmetry of Li$_2$(Pd$_{1-x}$,Pt$_x$)$_3$B, the SO coupling has
the form
\begin{equation}
\label{gamma_O}
    \bgam(\bk)=\gamma_0\bm{\phi}_{F_1,u}(\bk),
\end{equation}
where $\bm{\phi}_{F_1,u}$ are the odd basis functions of the
vector representation $F_1$, e.g.
$\bm{\phi}_{F_1,u}(\bk)\propto(k_x,k_y,k_z)$.

The Hamiltonian (\ref{H SO eff}) is diagonalized by a unitary
transformation:
\begin{equation}
\label{band transform}
    a_{\bk\alpha}=\sum_{\lambda=\pm}u_{\alpha\lambda}(\bk)c_{\bk\lambda},
\end{equation}
where
\begin{equation}
\label{Rashba_spinors}
    \begin{array}{l}
    \displaystyle u_{\uparrow\lambda}(\bk)=e^{i\theta_\lambda}
    \sqrt{\frac{|\bgam|+\lambda\gamma_z}{2|\bgam|}},\\
    \displaystyle u_{\downarrow\lambda}(\bk)=\lambda e^{i\theta_\lambda}
    \frac{\gamma_x+i\gamma_y}{\sqrt{2|\bgam|(|\bgam|+\lambda\gamma_z)}},
    \end{array}
\end{equation}
and $\theta_\lambda$ are arbitrary (in general $\bk$-dependent)
phases. The free-electron Hamiltonian then becomes
\begin{equation}
\label{H_0_band}
    H_0=\sum_{\bk}\sum_{\lambda=\pm}\xi_\lambda(\bk)c^\dagger_{\bk\lambda}c_{\bk\lambda},
\end{equation}
where
$\xi_\lambda(\bk)=\epsilon(\bk)+\lambda|\bgam(\bk)|=\xi_\lambda(-\bk)$
describes the quasiparticle dispersion in the $\lambda$th band.
Thus the SO coupling lifts the spin degeneracy of the electron
bands, so that the Fermi surface consists of two pieces defined by
the equations $\xi_\pm(\bk)=0$. The magnitude of the band
splitting is given by $2|\bgam(\bk)|$ and might vanish, for
symmetry reasons, along some directions or at some isolated points
in the Brillouin zone. We assume that the band structure is such
that the zeros of $\bgam(\bk)$ are not located on the Fermi
surface. This is the case in the model (\ref{gamma_O}), and also
in the model (\ref{gamma C4v}) if the Fermi surface is a cylinder
around the $z$-axis.

The Zeeman coupling of the electron spins with an external
magnetic field is described by
\begin{eqnarray}
\label{H Zeeman}
    H_Z&=&-\mu_B\bB\sum\limits_{\bk,\alpha\beta}
    \bm{\sigma}_{\alpha\beta}a^\dagger_{\bk\alpha}a_{\bk\beta}\nonumber\\
    &=&-\bB\sum_{\bk,\lambda\lambda'}\bm{m}_{\lambda\lambda'}(\bk)
    c^\dagger_{\bk\lambda}c_{\bk\lambda'},
\end{eqnarray}
where $\mu_B$ is the Bohr magneton. The components of
$\hat{\bm{m}}(\bk)=\mu_B\hat u^\dagger(\bk)\hat{\bm{\sigma}}\hat
u(\bk)$ have the following form:
\begin{eqnarray}
\label{m_i}
    &&\hat m_x=\mu_B
    \left(\begin{array}{cc}
      \hat\gamma_x & -e^{-i\tilde\theta}
      \frac{\gamma_x\hat\gamma_z+i\gamma_y}{\gamma_\perp} \\
      -e^{i\tilde\theta}
      \frac{\gamma_x\hat\gamma_z-i\gamma_y}{\gamma_\perp} & -\hat\gamma_x \\
    \end{array}\right),\nonumber\\
    &&\hat m_y=\mu_B
    \left(\begin{array}{cc}
      \hat\gamma_y & -e^{-i\tilde\theta}
      \frac{\gamma_y\hat\gamma_z-i\gamma_x}{\gamma_\perp} \\
      -e^{i\tilde\theta}
      \frac{\gamma_y\hat\gamma_z+i\gamma_x}{\gamma_\perp} & -\hat\gamma_y \\
    \end{array}\right),\quad\\
    &&\hat m_z=\mu_B
    \left(\begin{array}{cc}
      \hat\gamma_z & e^{-i\tilde\theta}\frac{\gamma_\perp}{\gamma} \\
      e^{i\tilde\theta}\frac{\gamma_\perp}{\gamma} & -\hat\gamma_z \\
    \end{array}\right)\nonumber,
\end{eqnarray}
where $\hat\bgam=\bgam/|\bgam|$,
$\gamma_\perp=\sqrt{\gamma_x^2+\gamma_y^2}$, and
$\tilde\theta=\theta_+-\theta_-$. The expectation value of the
spin magnetic moment of an electron from the $\lambda$th band with
the wave vector $\bk$ is
$\bm{m}_\lambda(\bk)=\lambda\mu_B\hat{\bgam}(\bk)$. We shall see
that, although the interband matrix elements of
$\hat{\bm{m}}(\bk)$ contain the arbitrary phases of the Bloch
spinors, those will not affect observable quantities.

Finally, we introduce the pairing interaction between electrons in
the Cooper channel, using the basis of the exact eigenstates of
the noninteracting problem:
\begin{eqnarray}
\label{H int}
    H_{int}=\frac{1}{2{\cal V}}\sum\limits_{\bk\bk'\bq}\sum_{\lambda\lambda'}
    V_{\lambda\lambda'}(\bk,\bk')c^\dagger_{\bk+\bq/2,\lambda}
    c^\dagger_{-\bk+\bq/2,\lambda}\nonumber\\
    \times c_{-\bk'+\bq/2,\lambda'}c_{\bk'+\bq/2,\lambda'},
\end{eqnarray}
where ${\cal V}$ is the system volume. The pairing potential
satisfies the relations
$V_{\lambda\lambda'}(-\bk,\bk')=-V_{\lambda\lambda'}(\bk,\bk')=V_{\lambda\lambda'}(\bk,-\bk')$,
which follow from the anti-commutation of the Fermi operators. The
diagonal elements of the matrix $\hat V$ describe the intraband
Cooper pairing, while the off-diagonal ones correspond to the pair
scattering from one band to the other. We assume, in the spirit of
the BCS theory, that the pairing interaction is nonzero only
inside the thin shells of width $\varepsilon_c$ in the vicinity of
the Fermi surfaces, i.e. when
$|\xi_{\lambda}(\bk)|,|\xi_{\lambda'}(\bk')|\leq\varepsilon_c$. We
further assume that it can be represented in a factorized form:
$V_{\lambda\lambda'}(\bk,\bk')=t_\lambda(\bk)t^*_{\lambda'}(\bk')
\tilde V_{\lambda\lambda'}(\bk,\bk')$, where
$t_\lambda(\bk)=-t_\lambda(-\bk)$ are nontrivial phase factors
which appear in the expression for the time-reversal operation for
nondegenerate bands:
$K|\bk\lambda\rangle=t_\lambda(\bk)|-\bk,\lambda\rangle$,\cite{GR01,SC04}
and
\begin{equation}
\label{pairing potential}
    \tilde V_{\lambda\lambda'}(\bk,\bk')=-V_{\lambda\lambda'}
    \sum_{a=1}^{d_\Gamma}\phi_{\lambda,a}(\bk)
    \phi^*_{\lambda',a}(\bk')
\end{equation}
is invariant under the point group operations: $\tilde
V_{\lambda\lambda'}(g^{-1}\bk,g^{-1}\bk')=\tilde
V_{\lambda\lambda'}(\bk,\bk')$. The coupling constants
$V_{\lambda\lambda'}$ form a symmetric positive-definite $2\times
2$ matrix, and $\phi_{\lambda,a}(\bk)$ are even basis functions of
an irreducible $d_\Gamma$-dimensional representation $\Gamma$ of
$\mathbb{G}$.\cite{Book} While $\phi_{+,a}(\bk)$ and
$\phi_{-,a}(\bk)$ have the same symmetry, their momentum
dependence does not have to be exactly the same. The basis
functions are nonzero only inside the BCS shells and are
normalized:
$\langle\phi^*_{\lambda,a}\phi_{\lambda,b}\rangle_\lambda=\delta_{ab}$,
where the angular brackets denote the averaging over the Fermi
surface in the $\lambda$th band.

Treating the pairing interaction (\ref{H int}) in the mean-field
approximation, one introduces the superconducting order
parameters, which in the uniform case considered here depend only
on $\bk$ and have the form
$\Delta_\lambda(\bk)=t_\lambda(\bk)\tilde\Delta_\lambda(\bk)$,
where $\tilde\Delta_\lambda(\bk)=\tilde\Delta_\lambda(-\bk)$ can
be represented as follows:
\begin{equation}
\label{OP definition}
    \tilde\Delta_\lambda(\bk)=\sum_a\eta_{\lambda,a}\phi_{\lambda,a}(\bk).
\end{equation}
Thus our system is formally equivalent to a two-band
superconductor, in which the order parameter has $2d_\Gamma$
components given by the expansion coefficients $\eta_{\lambda,a}$.
The order parameter components depend on temperature and vanish in
the normal state.

\subsection{Spin susceptibility}
\label{sec: chi clean}

Writing the Zeeman Hamiltonian (\ref{H Zeeman}) in the form
$H_Z=-{\cal{\bM}}\bB$, where ${\cal{\bM}}$ is the operator of the
total spin magnetic moment of electrons, we define the
magnetization as $\bM={\cal V}^{-1}\langle{\cal\bM}\rangle$ (the
angular brackets denote the thermodynamic average). In a weak
field, $M_i=\sum_j\chi_{ij}B_j$, where $\chi_{ij}$ is the spin
susceptibility tensor.

Introducing the four-component Nambu operators in the band
representation:
$C_{\bk}=(c_{\bk+},c_{\bk-},c^\dagger_{-\bk,+},c^\dagger_{-\bk,+})^T$,
we combine the normal and anomalous Green's functions\cite{AGD}
into a $4\times 4$ matrix Green's function
\begin{equation}
\label{matrix G def}
    {\cal G}(\bk_1,\bk_2;\tau)=-\langle T_\tau
    C_{\bk_1}(\tau)C^\dagger_{\bk_2}(0)\rangle.
\end{equation}
In the clean system, the Green's function is diagonal in momentum:
\begin{equation}
\label{matrix G}
    {\cal G}(\bk,\omega_n)=\left(\begin{array}{cc}
        \hat G(\bk,\omega_n) & -\hat F(\bk,\omega_n) \\
        -\hat F^\dagger(\bk,\omega_n) & -\hat G^T(-\bk,-\omega_n) \\
    \end{array}\right).
\end{equation}
Here $\omega_n=(2n+1)\pi T$ is the fermionic Matsubara frequency
(in our units $k_B=1$), and the hats here denote $2\times 2$
matrices in the band space. In the thermodynamic limit ${\cal
V}\to\infty$ the magnetization can be expressed in terms of the
Green's functions as follows:
\begin{equation}
\label{M via G}
    \bM=T\sum_n\int\frac{d^3\bk}{(2\pi)^3}\tr\,\hat{\bm{m}}(\bk)
    \hat G(\bk,\omega_n),
\end{equation}
where the matrices $\hat{\bm{m}}$ are given by Eqs. (\ref{m_i}).

In the presence of magnetic field we have ${\cal
G}(\bk,\omega_n)=[{\cal
G}^{-1}_0(\bk,\omega_n)-\Sigma_Z(\bk)]^{-1}$, where
\begin{equation}
\label{matrix G0}
    {\cal G}_0^{-1}(\bk,\omega_n)=\left(\begin{array}{cc}
    i\omega_n-\hat\xi(\bk) & -\hat\Delta(\bk) \\
    -\hat\Delta^\dagger(\bk) & i\omega_n+\hat\xi(\bk)\\
    \end{array}\right),
\end{equation}
$\Delta_{\lambda\lambda'}(\bk)=\delta_{\lambda\lambda'}t_\lambda(\bk)\tilde\Delta_\lambda(\bk)$,
and the Zeeman coupling is described by
\begin{equation}
\label{Sigma B}
    \Sigma_Z(\bk)=\left(\begin{array}{cc}
    -\hat{\bm{m}}(\bk)\bB & 0 \\
    0 & \hat{\bm{m}}^T(-\bk)\bB \\
    \end{array}\right).
\end{equation}
Expanding ${\cal G}$ in powers of $\bB$, we obtain:
\begin{equation}
\label{chi ij gen SC}
    \chi_{ij}=-T\sum_n\int\frac{d^3\bk}{(2\pi)^3}\tr\bigl(
    \hat m_i\hat G\hat m_j\hat G-\hat m_i\hat F\hat{\bar{m}}_j\hat
    F^\dagger\bigr),
\end{equation}
where $\hat{\bar{\bm{m}}}(\bk)=\hat{\bm{m}}^T(-\bk)$. The Green's
functions here are calculated at zero field and have the following
form:
\begin{eqnarray}
\label{GFs clean}
    &&G_{\lambda\lambda'}(\bk,\omega_n)=\delta_{\lambda\lambda'}G_\lambda(\bk,\omega_n),\nonumber\\
    &&F_{\lambda\lambda'}(\bk,\omega_n)=\delta_{\lambda\lambda'}t_\lambda(\bk)
    \tilde F_\lambda(\bk,\omega_n),\\
    &&F^\dagger_{\lambda\lambda'}(\bk,\omega_n)=\delta_{\lambda\lambda'}t^*_\lambda(\bk)
    \tilde F^*_\lambda(\bk,\omega_n),\nonumber
\end{eqnarray}
where
\begin{equation}
\label{gf clean}
    \begin{array}{l}
    \displaystyle G_\lambda(\bk,\omega_n)=-\frac{i\omega_n+\xi_\lambda(\bk)}{\omega_n^2
    +\xi_\lambda^2(\bk)+|\tilde\Delta_\lambda(\bk)|^2},\\ \\
    \displaystyle \tilde F_\lambda(\bk,\omega_n)=\frac{\tilde\Delta_\lambda(\bk)}{\omega_n^2
    +\xi_\lambda^2(\bk)+|\tilde\Delta_\lambda(\bk)|^2}. \\
    \end{array}
\end{equation}

Inserting the expressions (\ref{m_i}) for the electron magnetic
moment in Eq. (\ref{chi ij gen SC}) and using the identities
\begin{eqnarray}
\label{identity-4}
    &&\bar{m}_{i,\lambda\lambda'}=
        -t^*_\lambda t_{\lambda'}m_{i,\lambda\lambda'},\nonumber\\
    &&m_{i,++}m_{j,++}=m_{i,--}m_{j,--}=\mu_B^2\hat\gamma_i\hat\gamma_j,\nonumber\\
    &&m_{i,+-}m_{j,-+}=\mu_B^2(\delta_{ij}-\hat\gamma_i\hat\gamma_j
    +ie_{ijk}\hat\gamma_k),
\end{eqnarray}
we find that the susceptibility tensor can be represented in the
following form:
\begin{equation}
\label{chi ij composition}
    \chi_{ij}=\sum_{\lambda=\pm}\chi^{\lambda}_{ij}+\tilde\chi_{ij}.
\end{equation}
Here
\begin{eqnarray}
\label{chi intra gen SC}
    \chi^{\lambda}_{ij}&=&-\mu_B^2T\sum_n\int\frac{d^3\bk}{(2\pi)^3}
    \hat\gamma_i\hat\gamma_j\bigl(G_\lambda^2+|\tilde F_\lambda|^2\bigr)\nonumber\\
    &=&\mu_B^2N_\lambda\left\langle\hat\gamma_i\hat\gamma_jY_\lambda\right\rangle_\lambda
\end{eqnarray}
are the intraband contributions, which are determined by the
thermally-excited quasiparticles near the Fermi surfaces,
$N_\lambda$ is the density of states in the $\lambda$th band, and
\begin{equation}
\label{Y_lambda}
    Y_\lambda(\bk,T)=\frac{1}{2T}\int_0^\infty
   \frac{d\xi}{\cosh^2\bigl(\sqrt{\xi^2+|\tilde\Delta_\lambda(\bk)|^2}/2T\bigr)}
\end{equation}
is the angle-resolved Yosida function.

The last term in Eq. (\ref{chi ij composition}) is
\begin{eqnarray}
\label{chi inter gen SC}
    \tilde\chi_{ij}=-2\mu_B^2T\sum_n\int\frac{d^3\bk}{(2\pi)^3}
    (\delta_{ij}-\hat\gamma_i\hat\gamma_j)\nonumber\\
    \times\bigl(G_+G_-+\re\tilde F_+^*\tilde F_-\bigr).
\end{eqnarray}
In the normal state this becomes
\begin{equation}
\label{chi inter n}
    \tilde\chi_{ij}=-\mu_B^2\int\frac{d^3\bk}{(2\pi)^3}
    \frac{\delta_{ij}-\hat\gamma_i\hat\gamma_j}{|\bgam|}
    [f(\xi_+)-f(\xi_-)],
\end{equation}
where $f(\epsilon)=(e^{\epsilon/T}+1)^{-1}$ is the Fermi-Dirac
distribution function. In contrast to the intraband
susceptibilities $\chi^\lambda_{ij}$, which depend only on the
quasiparticle properties in the vicinity of the Fermi surfaces,
$\tilde\chi_{ij}$ is determined by all quasiparticles in the
momentum-space shell ``sandwiched'' between the Fermi surfaces. We
therefore call this the interband contribution (its physical
origin is discussed in Sec. \ref{sec: interband chi origin}
below). A straightforward calculation shows that even at $T=0$
$$
    \frac{\tilde\chi_{ij}|_{\tilde\Delta\neq 0}
    -\tilde\chi_{ij}|_{\tilde\Delta=0}}{\tilde\chi_{ij}|_{\tilde\Delta=0}}\sim
    \frac{|\tilde\Delta|^2}{|\bgam|^2}\ln\frac{|\bgam|}{|\tilde\Delta|}
    \ll 1,
$$
which means that $\tilde\chi_{ij}$ is almost unchanged when the
system undergoes a phase transition in which only the electrons
near the Fermi surface are affected.

Thus we arrive at the following expression for the spin
susceptibility of a clean superconductor:
\begin{equation}
\label{chi ij SC final}
    \chi_{ij}=\tilde\chi_{ij}+
    \mu_B^2N_F\sum_\lambda\rho_\lambda\left\langle\hat\gamma_i\hat\gamma_j
    Y_\lambda\right\rangle_\lambda,
\end{equation}
where $\tilde\chi_{ij}$ is given by Eq. (\ref{chi inter n}),
$N_F=(N_++N_-)/2$, and $\rho_\lambda=N_\lambda/N_F$. In the normal
state, $Y_\lambda=1$ and
\begin{equation}
\label{chi ij N}
    \chi_{N,ij}=\tilde\chi_{ij}+
    \mu_B^2N_F\sum_\lambda\rho_\lambda\left\langle\hat\gamma_i\hat\gamma_j
    \right\rangle_\lambda.
\end{equation}
At zero temperature, there is no excitations ($Y_\lambda=0$) and
the intraband contributions are absent, but the susceptibility
still attains a nonzero value given by $\tilde\chi_{ij}$. The
temperature dependence of the susceptibility in the
superconducting state at $0<T\leq T_c$ is almost entirely
determined by the intraband terms, with the low-temperature
behavior depending crucially on the magnitude of the SO coupling
at the gap nodes.\cite{Sam05} While in the fully gapped case the
intraband susceptibility is exponentially small in all directions,
in the presence of the lines of nodes it is proportional to either
$T$ or $T^3$, depending on whether or not the zeros of
$\tilde\Delta_\lambda(\bk)$ coincide with those of $\bgam(\bk)$,
see Ref. \onlinecite{Sam05} for details.

Further steps depend on the pairing symmetry and the electron
structure of the superconductor. Let us evaluate the expression
(\ref{chi ij SC final}) in the case when the SO coupling is small
compared to the Fermi energy, i.e. $|\bgam|\ll\epsilon_F$, and the
gaps in both bands are isotropic and have the same magnitude. It
is legitimate to neglect the difference between the densities of
states in the two bands: $\rho_+=\rho_-=1$, and we also have
$Y_+(\bk,T)=Y_-(\bk,T)=Y(T)$. The interband contribution (\ref{chi
inter n}) is reduced to
\begin{equation}
\label{tilde chi ij}
    \tilde\chi_{ij}=2\mu_B^2N_F(\delta_{ij}-
    \left\langle\hat\gamma_i\hat\gamma_j\right\rangle_F),
\end{equation}
where $\langle(...)\rangle_F$ denotes the average over the Fermi
surface defined by the equation $\epsilon(\bk)=0$. The
normal-state susceptibility is isotropic:
$\chi_{N,ij}=\chi_P\delta_{ij}$, where $\chi_P=2\mu_B^2N_F$ is the
Pauli susceptibility, while in the superconducting state
\begin{equation}
\label{chi ij gamma small}
    \chi_{ij}=\tilde\chi_{ij}
    +2\mu_B^2N_F\langle\hat\gamma_i\hat\gamma_j\rangle_F Y(T).
\end{equation}

\begin{figure}
    \includegraphics[width=\columnwidth]{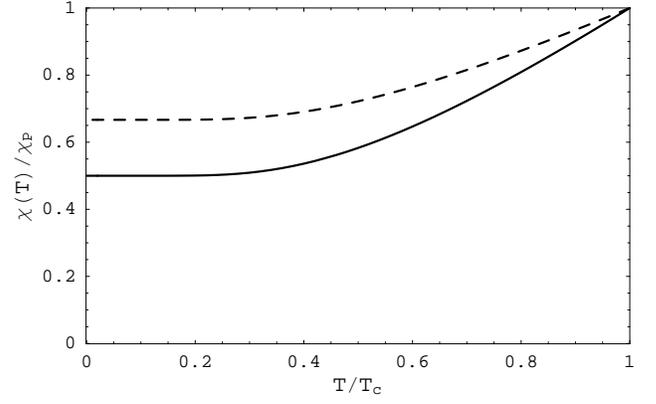}
    \caption{The temperature dependence of the transverse components of
    the susceptibility for the 2D model (the solid line), and of all
    three components for the 3D model (the dashed line). The $\chi_{zz}$ component
    in the 2D case is temperature-independent.}
    \label{fig: chi clean}
\end{figure}

Suppose the Fermi surface is a cylinder around the $z$ axis and
the SO coupling is described by the Rashba expression:
$\bgam(\bk)=\gamma_\perp(\bk\times\hat z)$. We will refer to this
as the two-dimensional (2D) model. The susceptibility tensor is
diagonal, with
\begin{equation}
\label{chi ij model I}
    \begin{array}{l}
    \displaystyle\chi_{xx}=\chi_{yy}=\frac{1}{2}[1+Y(T)]\chi_P,\qquad
    \displaystyle\chi_{zz}=\chi_P.
    \end{array}
\end{equation}
For a spherical Fermi surface in a cubic crystal with the SO
coupling described by $\bgam(\bk)=\gamma_0\bk$ [the
three-dimensional (3D) model], the susceptibility tensor is given
by
\begin{equation}
\label{chi ij model II}
    \chi_{xx}=\chi_{yy}=\chi_{zz}=\frac{1}{3}[2+Y(T)]\chi_P.
\end{equation}
The temperature dependence of $\chi_{ij}$ for the 2D and 3D models
is sketched in Fig. \ref{fig: chi clean}. In both cases the
susceptibility at $T=0$ is nonzero but still considerably less
than in the normal state.

\subsection{Origin of the residual susceptibility}
\label{sec: interband chi origin}

The mean-field pairing Hamiltonian in the band representation is
given by
$H_{sc}=(1/2)\sum_{\bk,\lambda}[\Delta_\lambda(\bk)c^\dagger_{\bk\lambda}
c^\dagger_{-\bk,\lambda}+\mathrm{H.c.}]$. This can be transformed
into the spin representation using Eqs. (\ref{band transform}):
$H_{sc}=(1/2)\sum_{\bk,\alpha\beta}[\Delta_{\alpha\beta}(\bk)a^\dagger_{\bk\alpha}
a^\dagger_{-\bk,\beta}+\mathrm{H.c.}]$, where
\begin{eqnarray}
\label{Delta spin}
    \Delta_{\alpha\beta}(\bk)=\sum_\lambda
    u_{\alpha\lambda}(\bk)\Delta_\lambda(\bk)u_{\beta\lambda}(-\bk)\nonumber\\
    =-[\hat u(\bk)\hat{\tilde\Delta}(\bk)\hat
    u^\dagger(\bk)(i\hat\sigma_2)]_{\alpha\beta}.
\end{eqnarray}
To obtain this, we used the identity
$u_{\beta\lambda}(-\bk)=t^*_\lambda(\bk)\sum_\gamma(i\sigma_2)_{\beta\gamma}
u^*_{\gamma\lambda}(\bk)$, which follows from the definition of
$t_\lambda(\bk)$. The next step is to write
$\hat{\tilde\Delta}=(\tilde\Delta_++\tilde\Delta_-)\hat\tau_0/2+
    (\tilde\Delta_+-\tilde\Delta_-)\hat\tau_3/2$, where
$\hat{\bm{\tau}}$ are the Pauli matrices. Inserting this in Eq.
(\ref{Delta spin}) and using
\begin{equation}
\label{identity-ut3u}
    \hat u(\bk)\hat\tau_3\hat u^\dagger(\bk)=\hat\bgam(\bk)\hat{\bm{\tau}},
\end{equation}
we obtain the order parameter in the spin representation as
follows:
\begin{equation}
\label{Delta spin final}
    \Delta_{\alpha\beta}(\bk)=\psi(\bk)(i\hat\sigma_2)_{\alpha\beta}+
    \bm{d}(\bk)(i\hat{\bm{\sigma}}\hat\sigma_2)_{\alpha\beta},
\end{equation}
where
\begin{equation}
\label{spin-singlet}
    \psi(\bk)=-\frac{\tilde\Delta_+(\bk)+\tilde\Delta_-(\bk)}{2}
\end{equation}
is the spin-singlet component, and
\begin{equation}
\label{spin-triplet}
    \bm{d}(\bk)=-\frac{\tilde\Delta_+(\bk)-\tilde\Delta_-(\bk)}{2}\hat\bgam(\bk)
\end{equation}
is the spin-triplet component.\cite{negative signs} In agreement
with the results of Ref. \onlinecite{FAKS04}, only one component
of the triplet order parameter survives (is ``protected'') in the
limit of a large SO band splitting.

We see that the residual susceptibility is not related to the
presence of the spin-triplet order parameter: Expression
(\ref{spin-triplet}) vanishes if the both gaps are the same, but
the residual susceptibility is still nonzero and given by Eq.
(\ref{chi inter n}). The origin of the residual susceptibility can
be understood using the following argument.\cite{Yip02} For a
given wave vector $\bk$, there are two electron states
$|\bk,\pm\rangle$ with energies $\xi_\pm(\bk)$, in which the
expectation values of the spin magnetic moment are
$\bm{m}_\pm(\bk)=\pm\mu_B\hat{\bgam}(\bk)$. Due to the presence of
the projector $\delta_{ij}-\hat\gamma_i\hat\gamma_j$, only the
component of $\bB$ which is perpendicular to $\bm{m}_\pm(\bk)$
contributes to the susceptibility (\ref{chi inter n}). Calculating
the energies of the states $|\bk,\pm\rangle$ in the second order
of the perturbation theory with the help of the identity
(\ref{identity-4}), we obtain:
$$
    \delta\xi^{(2)}_\lambda(\bk)=\lambda\mu_B^2\sum_{i,j}B_iB_j
    \frac{\delta_{ij}-\hat\gamma_i\hat\gamma_j}{2|\bgam|}.
$$
To find the total energy shift, we sum the contributions from all
$\bk$ and both bands, with the weights given by the Fermi-Dirac
distribution. In this way we recover Eq. (\ref{chi inter n}). Thus
the interband susceptibility appears, similarly to the Van Vleck
paramagnetism of atoms and solids, in the second order of the
perturbation theory, due to the field-induced virtual transitions
between the SO split electron bands.

\section{Effects of impurities}
\label{sec: impurities}

The effect of scalar impurities on the spin susceptibility in the
superconducting state is described by the Hamiltonian
$H=H_0+H_Z+H_{int}+H_{imp}$, where the first three terms are given
by Eqs. (\ref{H_0_band}), (\ref{H Zeeman}) and (\ref{H int}), and
\begin{equation}
\label{H_imp}
    H_{imp}=\int d^3\br\sum_\alpha
    U_{imp}(\br)\psi^\dagger_\alpha(\br)\psi_\alpha(\br).
\end{equation}
The disorder potential $U_{imp}(\br)$ is assumed to be a random
function with zero mean and the correlator $\langle
U_{imp}(\br_1)U_{imp}(\br_2)\rangle=n_{imp}U_0^2\delta(\br_1-\br_2)$,
where $n_{imp}$ is the impurity concentration, and $U_0$ is the
strength of an individual point-like impurity. The electron field
operators have the form
$$
    \psi_\alpha(\br)=\frac{1}{\sqrt{{\cal V}}}\sum_{\bk,\lambda}
    u_{\alpha\lambda}(\bk)e^{i\bk\br}c_{\bk\lambda},
$$
from which we obtain the band representation of the impurity
Hamiltonian:
\begin{equation}
\label{H_imp_band}
    H_{imp}=\frac{1}{{\cal V}}\sum_{\bk\bk'}\sum_{\lambda\lambda'}
    U_{imp}(\bk-\bk')w_{\lambda\lambda'}(\bk,\bk')
    c^\dagger_{\bk\lambda}c_{\bk'\lambda'}.
\end{equation}
Here $U_{imp}(\bq)$ is the Fourier transform of the impurity
potential, $\langle
U_{imp}(\bq)U_{imp}(\bq')\rangle=n_{imp}U_0^2{\cal
V}\delta_{\bq,-\bq'}$, and $\hat w(\bk,\bk')=\hat
u^\dagger(\bk)\hat u(\bk')=\hat w^\dagger(\bk',\bk)$. We see that
the impurity scattering amplitude in the band representation
acquires both intraband and interband contributions and also
becomes anisotropic, even for isotropic impurities.\cite{MS07}

Averaging with respect to the impurity positions restores
translational invariance: $\langle{\cal
G}(\bk_1,\bk_2;\omega_n)\rangle_{imp}=\delta_{\bk_1,\bk_2}{\cal
G}(\bk,\omega_n)$. The disorder-averaged Green's function here has
the same matrix structure as in the clean case, see Eq.
(\ref{matrix G}). The magnetization in the superconducting state
is determined by Eq. (\ref{M via G}), in which the Green's
function of the clean normal metal should be replaced by its
disorder average.

The average matrix Green's function satisfies the Gor'kov
equations: $({\cal G}_0^{-1}-\Sigma_{imp}-\Sigma_Z){\cal G}=1$,
where ${\cal G}_0$ is the Green's function at zero field in the
absence of impurities, given by Eq. (\ref{matrix G0}), the
impurity self-energy in the self-consistent Born approximation is
\begin{eqnarray}
\label{Sigma imp}
    &&\Sigma_{imp}(\bk,\omega_n)=n_{imp}U_0^2\nonumber\\
    &&\qquad\times\int\frac{d^3\bk'}{(2\pi)^3}W(\bk,\bk')
    {\cal G}(\bk',\omega_n)W(\bk',\bk),\qquad
\end{eqnarray}
and $\Sigma_Z$ is the Zeeman self-energy (\ref{Sigma B}). The
$4\times 4$ matrix $W$ is defined as follows:
$$
    W(\bk,\bk')=
    \left(\begin{array}{cc}
    \hat w(\bk,\bk') & 0 \\
    0 & -\hat w^T(-\bk',-\bk) \\
    \end{array}\right).
$$
It is straightforward to show that $[\hat
w^T(-\bk',-\bk)]_{\lambda\lambda'}=t^*_\lambda(\bk)t_{\lambda'}(\bk')
w_{\lambda\lambda'}(\bk,\bk')$.

\subsection{Zero-field solution}
\label{sec: impurities zero B}

In the absence of magnetic field, $\Sigma_Z=0$, and, since ${\cal
G}_0$ is band-diagonal, we seek the solution of the Gor'kov
equation in a band-diagonal form given by Eqs. (\ref{GFs clean}).
For consistency, we require that the Nambu matrix components of
the self-energy are also band-diagonal:
\begin{equation}
\label{Sigma imp structure}
    (\Sigma_{imp})^{ab}_{\lambda\lambda'}(\bk,\omega_n)=
    \delta_{\lambda\lambda'}\varrho^{ab}_\lambda(\bk)\sigma^{ab}_\lambda(\bk,\omega_n),
\end{equation}
where $a,b=1,2$ are the Nambu (particle-hole) indices, and
$\varrho^{11}_\lambda(\bk)=\varrho^{22}_\lambda(\bk)=1$,
$\varrho^{12}_\lambda(\bk)=t_\lambda(\bk)$,
$\varrho^{21}_\lambda(\bk)=t^*_\lambda(\bk)$. From Eq. (\ref{Sigma
imp}) we obtain:
\begin{eqnarray}
    &&\left(\begin{array}{cc}
    \hat{\sigma}^{11}(\bk,\omega_n) & \hat{\sigma}^{12}(\bk,\omega_n) \\
    \hat{\sigma}^{21}(\bk,\omega_n) & \hat{\sigma}^{22}(\bk,\omega_n) \\
    \end{array}\right)\nonumber\\
    &&\quad=n_{imp}U_0^2\int\frac{d^3\bk'}{(2\pi)^3}\hat w(\bk,\bk')\nonumber\\
    &&\quad\times\left(\begin{array}{cc}
    \hat{G}(\bk',\omega_n) & \hat{\tilde F}(\bk',\omega_n) \\
    \hat{\tilde F}^\dagger(\bk',\omega_n) & -\hat{G}^T(-\bk',-\omega_n) \\
    \end{array}\right)\hat w(\bk',\bk).\quad
\end{eqnarray}
Using the identity (\ref{identity-ut3u}) one can show that the
integrands on the right-hand side of these equations have the
following form:
\begin{eqnarray}
\label{UGU}
    &&\hat u(\bk')\hat{G}(\bk',\omega_n)\hat u^\dagger(\bk')\nonumber\\
    &&\hspace*{1.5cm}=\frac{G_+(\bk',\omega_n)+G_-(\bk',\omega_n)}{2}\hat\tau_0\nonumber\\
    &&\hspace*{1.5cm}+\frac{G_+(\bk',\omega_n)-
    G_-(\bk',\omega_n)}{2}\hat\bgam(\bk')\hat{\bm{\tau}},\qquad
\end{eqnarray}
\emph{etc}. Assuming that $G_\lambda$ and $\tilde F_\lambda$ are
even functions of momentum (the self-consistency of this will be
verified below), the last line in Eq. (\ref{UGU}) is odd in $\bk'$
and therefore vanishes after the $\bk'$-integration. Then,
\begin{eqnarray*}
    &&\sigma^{11}_\lambda(\bk,\omega_n)=-\sigma^{22}_\lambda(\bk,-\omega_n)=\Sigma_1(\omega_n),\\
    &&\sigma^{12}_\lambda(\bk,\omega_n)=\sigma^{21,*}_\lambda(\bk,\omega_n)=\Sigma_2(\omega_n),
\end{eqnarray*}
where
\begin{equation}
\label{Sigma-12}
    \begin{array}{l}
    \displaystyle\Sigma_1(\omega_n)=\frac{1}{2}n_{imp}U_0^2\sum_\lambda\int\frac{d^3\bk}{(2\pi)^3}
    G_\lambda(\bk,\omega_n), \\
    \\
    \displaystyle\Sigma_2(\omega_n)=\frac{1}{2}n_{imp}U_0^2\sum_\lambda\int\frac{d^3\bk}{(2\pi)^3}
    {\tilde F}_\lambda(\bk,\omega_n).
    \end{array}
\end{equation}
Absorbing the real part of $\Sigma$ into the chemical potential,
we have $\Sigma_1(\omega_n)=i\tilde\Sigma_1(\omega_n)$, where
$\tilde\Sigma_1$ is odd in $\omega_n$.

Solving the Gor'kov equations we obtain the disorder-averaged
Green's functions:
\begin{equation}
\label{GFs average}
    \begin{array}{l}
    \displaystyle G_\lambda(\bk,\omega_n)=-\frac{i\tilde\omega_n+\xi_\lambda(\bk)}{\tilde\omega_n^2
    +\xi_\lambda^2(\bk)+|D_\lambda(\bk,\omega_n)|^2},\\ \\
    \displaystyle \tilde F_\lambda(\bk,\omega_n)=\frac{D_\lambda(\bk,\omega_n)}{\tilde\omega_n^2
    +\xi_\lambda^2(\bk)+|D(\bk,\omega_n)|^2},
    \end{array}
\end{equation}
where $\tilde\omega_n=\omega_n-\tilde\Sigma_1(\omega_n)$ and
$D_\lambda(\bk,\omega_n)=\tilde\Delta_\lambda(\bk)+\Sigma_2(\omega_n)$.
Substituting these expressions into Eqs. (\ref{Sigma-12}), we
arrive at the self-consistency equations for the renormalized
Matsubara frequency and the gap functions:
\begin{eqnarray}
\label{tilde omega eq}
    &&\tilde\omega_n=\omega_n+\frac{\Gamma}{2}\sum_\lambda\rho_\lambda
    \left\langle\frac{\tilde\omega_n}{\sqrt{\tilde\omega_n^2+|D_\lambda(\bk,\omega_n)|^2}}
    \right\rangle_\lambda,\qquad\\
\label{D eq}
    &&D_\lambda(\bk,\omega_n)=\tilde\Delta_\lambda(\bk)\nonumber\\
    &&\hspace*{1.45cm}+\frac{\Gamma}{2}\sum_{\lambda'}\rho_{\lambda'}
    \left\langle\frac{D_{\lambda'}(\bk,\omega_n)}{\sqrt{\tilde\omega_n^2
    +|D_{\lambda'}(\bk,\omega_n)|^2}}\right\rangle_{\lambda'}.
\end{eqnarray}
Here $\Gamma=1/2\tau$ is the elastic scattering rate, and
$\tau=(2\pi n_{imp}U_0^2N_F)^{-1}$ is the electron mean free time
due to impurities. One can now see that, while $\tilde\omega_n$ is
odd in $\omega_n$, $D_\lambda$ are even in both $\bk$ and
$\omega_n$, so that our assumptions are self-consistent.

\subsection{Spin susceptibility}
\label{sec: chi impurities}

At $\bB\neq 0$, both the Green's function and the impurity
self-energy acquire field-dependent corrections: ${\cal
G}_{\bB\neq 0}={\cal G}+\delta{\cal G}$, $\Sigma_{imp,\bB\neq
0}=\Sigma_{imp}+\delta\Sigma_{imp}$. Here ${\cal G}$ and
$\Sigma_{imp}$ are the average Green's function and the impurity
self-energy at zero field found in Sec. \ref{sec: impurities zero
B}. Treating the Zeeman coupling as a small perturbation, we find
$\delta{\cal G}={\cal G}(\Sigma_Z+\delta\Sigma_{imp}){\cal G}$.

Magnetization $\bM$ is given by Eq. (\ref{M via G}). The
contribution from the zero-field Green's function vanishes after
the momentum integration, meaning that there is no spontaneous
magnetism of the Cooper pairs in the superconducting state. Thus
the magnetization is determined by $\delta\hat G$ and can be
written as $\bM=\bM_1+\bM_2$, where
\begin{equation}
\label{M1}
    \bM_1=T\sum_n\int\frac{d^3\bk}{(2\pi)^3}\tr\,\hat{\bm{m}}({\cal G}
    \Sigma_Z{\cal G})^{11}
\end{equation}
corresponds diagrammatically to a ``bubble'' containing two
disorder-averaged Green's functions, while
\begin{equation}
\label{M2}
    \bM_2=T\sum_n\int\frac{d^3\bk}{(2\pi)^3}\tr\,\hat{\bm{m}}({\cal G}
    \delta\Sigma_{imp}{\cal G})^{11}
\end{equation}
describes the impurity vertex corrections (recall that the upper
indices here label the Nambu matrix components).

The contribution to the susceptibility from $\bM_1$ can be
calculated similarly to the clean case in Sec. \ref{sec: chi
clean}: $\chi_{1,ij}=\partial M_{1,i}/\partial
B_j=\chi^+_{ij}+\chi^-_{ij}+\tilde\chi_{ij}$, where
$\chi^{\lambda}_{ij}$ are the intraband and $\tilde\chi_{ij}$ the
interband susceptibilities. Since the latter is determined by all
quasiparticles between the ``$+$'' and ``$-$'' Fermi surfaces, its
change below the superconducting transition is negligibly small.
One can show that the interband susceptibility in the normal state
is not sensitive to impurities, therefore $\tilde\chi_{ij}$ is
given by its clean normal state expression (\ref{chi inter n}).
The intraband susceptibilities are given by
\begin{eqnarray}
\label{chi intra SC imp}
    \chi^{\lambda}_{ij}&=&-\mu_B^2T\sum_n\int\frac{d^3\bk}{(2\pi)^3}
    \hat\gamma_i\hat\gamma_j\bigl(G_\lambda^2+|\tilde F_\lambda|^2\bigr)\nonumber\\
    &=&\mu_B^2N_\lambda\left\langle\hat\gamma_i\hat\gamma_j\right\rangle_\lambda\nonumber\\
    &&-\pi\mu_B^2N_\lambda T\sum_n\left\langle\hat\gamma_i\hat\gamma_j
    \frac{|D_\lambda|^2}{(\tilde\omega_n^2+|D_\lambda|^2)^{3/2}}\right\rangle_\lambda,\quad
\end{eqnarray}
The Green's functions in the first line here are given by the
disorder-averaged expressions (\ref{GFs average}). In contrast to
the clean case, however, it is not possible to calculate the
Matsubara sums before the momentum integrals. In order to do the
momentum integrals first, one should\cite{AGD} add and subtract
the normal-state intraband susceptibility, which is not affected
by impurities, see Eq. (\ref{chi ij N}).

Let us now calculate the impurity vertex corrections (\ref{M2}).
Substituting $\delta{\cal G}={\cal
G}(\Sigma_Z+\delta\Sigma_{imp}){\cal G}$ into Eq. (\ref{Sigma
imp}), we obtain the following equation for $\delta\Sigma_{imp}$:
\begin{eqnarray}
\label{delta Sigma imp eq}
    \delta\Sigma_{imp}(\bk,\omega_n)-n_{imp}U_0^2\int\frac{d^3\bk'}{(2\pi)^3}W(\bk,\bk'){\cal G}(\bk',\omega_n)\nonumber\\
    \times\delta\Sigma_{imp}(\bk',\omega_n)
    {\cal G}(\bk',\omega_n)W(\bk',\bk)\nonumber\\
    =n_{imp}U_0^2\int\frac{d^3\bk'}{(2\pi)^3}W(\bk,\bk'){\cal G}(\bk',\omega_n)\nonumber\\
    \times\Sigma_Z(\bk'){\cal G}(\bk',\omega_n)W(\bk',\bk).
\end{eqnarray}
The contributions to $\bm{M}_2$ from the interband components of
$\delta\Sigma_{imp}$ can be neglected compared to those from the
intraband components, because the former contain the momentum
integrals of the products of the Green's functions from different
bands, which are at least by a factor of
$|\tilde\Delta|/|\bgam|\ll 1$ smaller than their same-band
counterparts. One can seek the field-induced correction to the
impurity self-energy in the band-diagonal form similar to Eq.
(\ref{Sigma imp structure}):
\begin{equation}
\label{delta Sigma imp structure}
    (\delta\Sigma_{imp})^{ab}_{\lambda\lambda'}(\bk,\omega_n)=
    \delta_{\lambda\lambda'}\varrho^{ab}_\lambda(\bk)\delta\sigma^{ab}_\lambda(\bk,\omega_n).
\end{equation}
For the same reason, one should retain only those terms on the
right-hand side of Eq. (\ref{delta Sigma imp eq}) in which both
Green's functions have the same band index. Since
$\bm{m}_{\lambda\lambda}(-\bk)=-\bm{m}_{\lambda\lambda}(\bk)=-\lambda\mu_B\hat\bgam(\bk)$,
see Eqs. (\ref{m_i}), we have
$(\Sigma_Z)^{11}_{\lambda\lambda}(\bk)=(\Sigma_Z)^{22}_{\lambda\lambda}(\bk)=-\lambda\mu_B\hat\bgam(\bk)$.
Inserting this and the expressions (\ref{delta Sigma imp
structure}) in Eq. (\ref{delta Sigma imp eq}) and using the
identity
$$
    |w_{\lambda\lambda'}(\bk,\bk')|^2=
    \frac{1+\lambda\lambda'\hat\bgam(\bk)\hat\bgam(\bk')}{2},
$$
we obtain the Nambu matrix components of the right-hand side of
Eq. (\ref{delta Sigma imp eq}):
\begin{eqnarray*}
    (R.H.S.)^{ab}_{\lambda}(\bk,\omega_n)=-\lambda\mu_B\varrho^{ab}_\lambda(\bk)
    \sum_{ij}\hat\gamma_i(\bk)B_j\nonumber\\
    \times\frac{1}{2}n_{imp}U_0^2\sum_{\lambda'}\int\frac{d^3\bk'}{(2\pi)^3}
    \hat\gamma_i\hat\gamma_jR_{\lambda'}^{ab},
\end{eqnarray*}
where $R^{11}_\lambda=G^2_{\lambda}+|\tilde F_{\lambda}|^2$,
$R^{12}_\lambda=(G_{\lambda}-\bar G_{\lambda})\tilde F_{\lambda}$,
$R^{21}_\lambda=(G_{\lambda}-\bar G_{\lambda})\tilde
F^*_{\lambda}$, $R^{22}_\lambda=\bar G^2_{\lambda}+|\tilde
F_{\lambda}|^2$, and $\bar
G_{\lambda}(\bk,\omega_n)=G_{\lambda}(-\bk,-\omega_n)$. This
suggests that one can seek $\delta\sigma^{ab}_\lambda$ in the
following form:
\begin{equation}
\label{delta sigma ab}
    \begin{array}{l}
    \delta\sigma^{11}_\lambda(\bk,\omega_n)=\delta\sigma^{22}_\lambda(\bk,\omega_n)=
    \lambda\hat\bgam(\bk)\bm{X}(\omega_n), \medskip \\
    \delta\sigma^{12}_\lambda(\bk,\omega_n)=-\delta\sigma^{21,*}_\lambda(\bk,\omega_n)=
    \lambda\hat\bgam(\bk)\bm{Y}(\omega_n).
    \end{array}
\end{equation}

After some straightforward algebra, we obtain from Eq. (\ref{delta
Sigma imp eq}) the linear equations for $\bm{X}(\omega_n)$ and
$\bm{Y}(\omega_n)$:
\begin{equation}
\label{XY eqs gen}
    \begin{array}{l}
    \displaystyle X_i-\sum_{j}(A_{1,ij}X_j+A_{2,ij}Y_j+A^*_{2,ij}Y^*_j)
    =X_{0,i}, \\
    \displaystyle Y_i-\sum_{j}(2A^*_{2,ij}X_j+A_{3,ij}Y_j+A_{4,ij}Y^*_j)
    =Y_{0,i}. \\
    \end{array}
\end{equation}
The notations here are as follows:
\begin{eqnarray*}
    &&A_{1,ij}(\omega_n)=
    \frac{\Gamma}{2}\sum_\lambda\rho_\lambda\left\langle\hat\gamma_i\hat\gamma_j
    \frac{|D_\lambda|^2}{(\tilde\omega_n^2+|D_\lambda|^2)^{3/2}}\right\rangle_\lambda,\\
    &&A_{2,ij}(\omega_n)=
    \frac{\Gamma}{4}\sum_\lambda\rho_\lambda\left\langle\hat\gamma_i\hat\gamma_j
    \frac{i\tilde\omega_nD_\lambda^*}{(\tilde\omega_n^2+|D_\lambda|^2)^{3/2}}\right\rangle_\lambda,\\
    &&A_{3,ij}(\omega_n)=
    \frac{\Gamma}{4}\sum_\lambda\rho_\lambda\left\langle\hat\gamma_i\hat\gamma_j
    \frac{2\tilde\omega_n^2+|D_\lambda|^2}{(\tilde\omega_n^2+|D_\lambda|^2)^{3/2}}\right\rangle_\lambda,\\
    &&A_{4,ij}(\omega_n)=
    \frac{\Gamma}{4}\sum_\lambda\rho_\lambda\left\langle\hat\gamma_i\hat\gamma_j
    \frac{D_\lambda^2}{(\tilde\omega_n^2+|D_\lambda|^2)^{3/2}}\right\rangle_\lambda,\\
    &&X_{0,i}(\omega_n)=-\mu_B\sum_jA_{1,ij}(\omega_n)B_j,\\
    &&Y_{0,i}(\omega_n)=-2\mu_B\sum_jA^*_{2,ij}(\omega_n)B_j.
\end{eqnarray*}
We see that, while $\bm{X}(\omega_n)$ is real, $\bm{Y}(\omega_n)$
is complex, and both are linear functions of $\bB$. For arbitrary
complex gap functions $\tilde\Delta_\lambda(\bk)$, Eqs. (\ref{XY
eqs gen}) can be transformed into a system of nine equations for
the components of $\bm{X}(\omega_n)$, $\re\bm{Y}(\omega_n)$, and
$\im\bm{Y}(\omega_n)$.

Plugging the expressions (\ref{delta Sigma imp structure}) and
(\ref{delta sigma ab}) into Eq. (\ref{M2}), we obtain:
\begin{eqnarray}
\label{M2 final}
    M_{2,i}=\mu_B T\sum_n\sum_\lambda\int\frac{d^3\bk}{(2\pi)^3}
    \hat\gamma_i\hat\gamma_j\bigl[(G_\lambda^2+|\tilde F_\lambda|^2)X_j\nonumber\\
    -(G_\lambda\tilde F_\lambda^*)Y_j
    +(G_\lambda\tilde F_\lambda)Y^*_j\bigr]\nonumber\\
    =4\pi\mu_BN_F\tau T\sum_n[X_i(\omega_n)-X_{0,i}(\omega_n)]
\end{eqnarray}
[here we also used the first of the equations (\ref{XY eqs gen})].
The vertex correction to the susceptibility is then found from
$\chi_{2,ij}=\partial M_{2,i}/\partial B_j$.

The total spin susceptibility has the form
$\chi_{ij}=\chi_{1,ij}+\chi_{2,ij}=\tilde\chi_{ij}+\chi^+_{ij}+\chi^-_{ij}+\chi_{2,ij}$.
Putting together Eqs. (\ref{chi intra SC imp}) and (\ref{M2
final}), we arrive at the following final expression:
\begin{equation}
\label{chi ij impurities gen}
    \chi_{ij}=\chi_{N,ij}+4\pi\mu_BN_F\tau T\sum_n
    \frac{\partial X_i(\omega_n)}{\partial B_j},
\end{equation}
where $\chi_{N,ij}$ is the normal-state susceptibility (\ref{chi
ij N}), which is not sensitive to disorder. At $\Gamma\to 0$, we
have $\bm{X}(\omega_n)\to\bm{X}_0(\omega_n)$, with
$\tilde\omega_n=\omega_n$ and $D_\lambda=\tilde\Delta_\lambda$.
Using the identity
$$
    \pi T\sum_n\frac{|\tilde\Delta_\lambda(\bk)|^2}{(\omega_n^2
    +|\tilde\Delta_\lambda(\bk)|^2)^{3/2}}=1-Y_\lambda(\bk,T),
$$
where $Y_\lambda$ is given by Eq. (\ref{Y_lambda}), we recover the
susceptibility (\ref{chi ij SC final}) of a clean superconductor.

\section{Residual susceptibility}
\label{sec: residual chi}

The equations derived in the previous section cannot be solved
analytically in the general case. On the other hand, application
of our results to real noncentrosymmetric materials is complicated
by the lack of a definite information about the superconducting
gap symmetry and the distribution of the pairing strength between
the bands. Therefore one can only make progress by using some
simple models.

First, we assume that the pairing corresponds to the unity
representation of the point group and is fully isotropic:
$\phi_+(\bk)=\phi_-(\bk)=1$, i.e.
$\tilde\Delta_\lambda(\bk)=\eta_\lambda$. Then it follows from
Eqs. (\ref{D eq}) that $D_\lambda(\bk,\omega_n)=D(\omega_n)$ and
the order parameter components (which can be chosen to be real)
satisfy the self-consistency equations:
\begin{eqnarray}
\label{gap eqs disorder}
    \sum_{\lambda'}V^{-1}_{\lambda\lambda'}\eta_{\lambda'}
    =T\sum_n\int\frac{d^3\bk}{(2\pi)^3}\tilde F_\lambda(\bk,\omega_n)\nonumber\\
    =\pi\rho_\lambda N_F T\sum_n
    \frac{D_\lambda(\omega_n)}{\sqrt{\tilde\omega_n^2+D^2_\lambda(\omega_n)}},
\end{eqnarray}
where $\tilde\omega_n$ is found from Eq. (\ref{tilde omega eq}).
The Matsubara sum is cut off at the limiting frequency of the
order of the BCS shell width $\varepsilon_c$.

We further assume that the SO coupling is weak compared to the
Fermi energy, $|\bgam|/\epsilon_F\ll 1$, and that the pairing
strength, see Eq. (\ref{pairing potential}) does not vary between
the bands: $V_{++}=V_{--}>0$. Since the matrix of the coupling
constants is symmetric and positive-definite, we also have
$V_{+-}=V_{-+}$ and $V_{++}>|V_{+-}|$. The self-consistency
equations have two solutions: 1) the order parameter magnitudes
and phases in the two bands are the same: $\eta_+=\eta_-=\eta$,
and 2) the magnitudes are the same, but the signs are opposite:
$\eta_+=-\eta_-=\eta$. According to Sec. \ref{sec: interband chi
origin}, the former solution corresponds to the singlet state in
the spin representation, while the latter -- to the ``protected''
triplet state. While in the clean limit the spin susceptibility
for both states is given by Eq. (\ref{chi ij gamma small}), the
effects of impurities in the two cases have to be analyzed
separately.

\subsection{$\eta_+=\eta_-=\eta$}
\label{sec: singlet}

In this case $D_+=D_-=D(\omega_n)$, and Eqs. (\ref{tilde omega
eq}) and (\ref{D eq}) take the following form:
\begin{eqnarray*}
    &&D=\eta+\Gamma\frac{D}{\sqrt{\tilde\omega_n^2+D^2}},\\
    &&\tilde\omega_n=\omega_n+\Gamma
    \frac{\tilde\omega_n}{\sqrt{\tilde\omega_n^2+D^2}}.
\end{eqnarray*}
The solution of these equations is $D(\omega_n)=Z(\omega_n)\eta$,
$\tilde\omega_n=Z(\omega_n)\omega_n$, where
$Z(\omega_n)=1+\Gamma/\sqrt{\omega_n^2+\eta^2}$. The gap equation
(\ref{gap eqs disorder}) becomes
\begin{equation}
\label{gap eq singlet}
    \eta=\pi g_1T\sum_n\frac{D}{\sqrt{\tilde\omega_n^2+D^2}}=
    \pi g_1T\sum_n\frac{\eta}{\sqrt{\omega_n^2+\eta^2}},
\end{equation}
where $g_1=(V_{++}+V_{+-})N_F$ is the dimensionless coupling
constant. Note that the scattering rate has dropped out of the gap
equation, so that there is an analog of the Anderson theorem:
neither the gap magnitude nor the critical temperature are
affected by impurities:
$T_c(\Gamma)=T_{c1}=(2\gamma\epsilon_c/\pi)e^{-1/g_1}$
($\ln\gamma\simeq 0.577$ is Euler's constant). In particular, the
gap magnitude at $T=0$ is given by the clean BCS expression:
$\eta(T=0)=\eta_0=(\pi/\gamma)T_{c1}$.

Let us now calculate the spin susceptibility, using Eq. (\ref{chi
ij impurities gen}). Since the gap functions are real, $\bm{Y}$ is
purely imaginary: $\bm{Y}(\omega_n)=i\tilde{\bm{Y}}(\omega_n)$,
and Eqs. (\ref{XY eqs gen}) take the following form:
\begin{equation}
\label{XY eqs singlet}
    \left(\begin{array}{cc}
    1-\hat C_1 & \hat C_2 \\
    \hat C_2 & 1-\hat C_3
    \end{array}\right)
    \left(\begin{array}{c}
    \bm{X} \\
    \tilde{\bm{Y}}
    \end{array}\right)=
    \left(\begin{array}{c}
    -\mu_B\hat C_1\bB \\
    \mu_B\hat C_2\bB
    \end{array}\right),
\end{equation}
where $C_{m,ij}=C_m\langle\hat\gamma_i\hat\gamma_j\rangle_F$,
$m=1,2,3$, and
\begin{eqnarray*}
    &&C_1(\omega_n)=\Gamma\frac{1}{Z(\omega_n)}
    \frac{\eta^2}{(\omega_n^2+\eta^2)^{3/2}},\\
    &&C_2(\omega_n)=\Gamma\frac{1}{Z(\omega_n)}
    \frac{\omega_n\eta}{(\omega_n^2+\eta^2)^{3/2}},\\
    &&C_3(\omega_n)=\Gamma\frac{1}{Z(\omega_n)}
    \frac{\omega_n^2}{(\omega_n^2+\eta^2)^{3/2}}.
\end{eqnarray*}
Substituting the solution of the equations (\ref{XY eqs singlet})
in Eq. (\ref{chi ij impurities gen}), we obtain:
\begin{equation}
\label{chi ij impurities singlet}
    \chi_{ij}=\chi_P\delta_{ij}-\pi\mu_B^2N_FT\sum_n\Lambda_{ij}(\omega_n),
\end{equation}
where $\hat\Lambda=4\tau(1-\hat{L})^{-1}\hat{L}$, and $\hat L=\hat
C_1+\hat C_2(1-\hat C_3)^{-1}\hat C_2$. In the coordinate system
in which $\langle\hat\gamma_i\hat\gamma_j\rangle_F$ is diagonal,
the susceptibility tensor is also diagonal:
\begin{equation}
    \frac{\chi_{ii}(T)}{\chi_P}=1-\langle\hat\gamma_i^2\rangle_F
    \pi T\sum_n\frac{\eta^2}{\omega_n^2+\eta^2}\frac{1}{\sqrt{\omega_n^2+\eta^2}
    +\Gamma_i},
\end{equation}
where $\Gamma_i=(1-\langle\hat\gamma_i^2\rangle_F)\Gamma$.

\begin{figure}
    \includegraphics[width=\columnwidth]{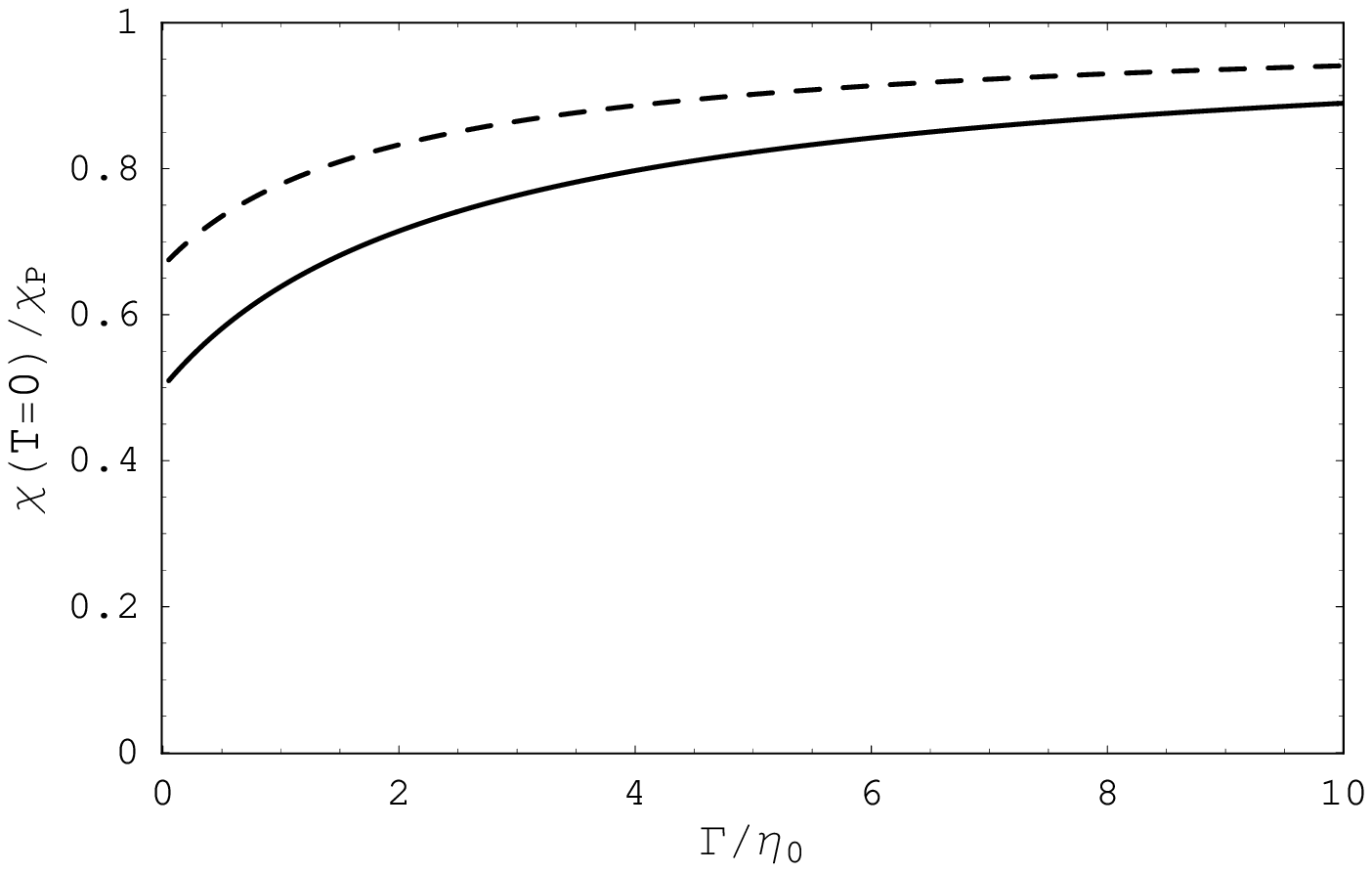}
    \caption{The residual susceptibility vs disorder strength for $\eta_+=\eta_-=\eta_0$.
    The solid line corresponds to the transverse components in the
    2D case ($\chi_{zz}=\chi_P$ and is disorder-independent), the dashed line
    -- to all three diagonal components in the 3D case.}
    \label{fig: chi_s}
\end{figure}

We are particularly interested in the effect of disorder on the
residual susceptibility at $T=0$. In this limit, the Matsubara sum
in the last formula can be replaced by a frequency integral, which
gives
\begin{equation}
\label{chi zero T singlet}
    \frac{\chi_{ii}(T=0)}{\chi_P}=
    1-\langle\hat\gamma_i^2\rangle_F+\langle\hat\gamma_i^2\rangle_F
    \Phi_1\left(\frac{\Gamma_i}{\eta_0}\right),
\end{equation}
where
\begin{equation}
\label{Theta1}
    \Phi_1(x)=1-\frac{\pi}{2x}\left(1-\frac{4}{\pi\sqrt{1-x^2}}\arctan\sqrt{\frac{1-x}{1+x}}\right)
\end{equation}
[at $x>1$ this function is evaluated using
$\arctan(ix)=i\tanh^{-1}(x)$]. While the first two terms on the
right-hand side of the expression (\ref{chi zero T singlet})
represent the residual susceptibility in the clean case, see Eq.
(\ref{tilde chi ij}), the last term describes the impurity effect.
In a weakly-disordered superconductor, using the asymptotics
$\Phi_1(x)\simeq\pi x/4$, we find that the residual susceptibility
increases linearly with disorder. In the dirty limit,
$\Gamma\gg\eta_0$, we have $\Phi_1(x)\to 1$, therefore
$\chi_{ii}(T=0)$ approaches the normal-state value $\chi_P$. For
the two simple band-structure models discussed in the end of Sec.
\ref{sec: chi clean} the Fermi-surface averages can be calculated
analytically, and we obtain the results plotted in Fig. \ref{fig:
chi_s}.

Thus we see that, similarly to spin-orbit impurities in a usual
centrosymmetric superconductor,\cite{AG62} scalar impurities in a
noncentrosymmetric superconductor lead to an enhancement of the
spin susceptibility at $T=0$. Since the interband contribution is
not sensitive to disorder, this effect can be attributed to an
increase in the intraband susceptibilities.

\subsection{$\eta_+=-\eta_-=\eta$}
\label{sec: triplet}

In this case $D_+=-D_-=\eta$, and we obtain from Eqs. (\ref{tilde
omega eq}) and (\ref{gap eqs disorder}):
\begin{eqnarray}
\label{tilde omega eq triplet}
    &&\tilde\omega_n=\omega_n+\Gamma
    \frac{\tilde\omega_n}{\sqrt{\tilde\omega_n^2+\eta^2}},\\
\label{gap eq triplet}
    &&\eta=\pi g_2 T\sum_n\frac{\eta}{\sqrt{\tilde\omega_n^2+\eta^2}},
\end{eqnarray}
where $g_2=(V_{++}-V_{+-})N_F$. In the absence of impurities, the
critical temperature is given by the BCS expression:
$T_c(\Gamma=0)=T_{c2}=(2\gamma\epsilon_c/\pi)e^{-1/g_2}$. If
$V_{+-}>0$ (attractive interband interaction), then $g_2<g_1$ and
$T_{c2}<T_{c1}$, i.e. the phase transition occurs into the state
$\eta_+=\eta_-$. If $V_{+-}<0$ (repulsive interband interaction),
then $g_2>g_1$ and $T_{c2}>T_{c1}$, i.e. the phase transition
occurs into the state $\eta_+=-\eta_-$.

In contrast to the previous case, both the critical temperature
and the gap magnitude are now suppressed by disorder. Indeed,
linearizing the equations (\ref{tilde omega eq triplet}) and
(\ref{gap eq triplet}) with respect to $\eta$, we obtain the
following equation for the critical temperature $T_c(\Gamma)$:
\begin{equation}
\label{Tc eq}
    \ln\frac{T_{c2}}{T_c}=\Psi\left(\frac{1}{2}+\frac{\Gamma}{2\pi T_c}\right)
    -\Psi\left(\frac{1}{2}\right).
\end{equation}
Thus the suppression of $T_c$ by scalar impurities is described by
the same Abrikosov-Gor'kov function as in a conventional BCS
superconductor with magnetic impurities. The superconductivity is
completely destroyed if the disorder strength exceeds the critical
value $\Gamma_c=(\pi/2\gamma)T_{c2}$.

To find the gap magnitude at $T=0$ we follow the procedure
outlined in Ref. \onlinecite{MK01}. Replacing the Matsubara sum by
a frequency integral in the gap equation (\ref{gap eq triplet}),
we have
\begin{eqnarray}
\label{eta eq triplet}
    \frac{1}{g_2}=\int_0^\infty d\omega\left(\frac{1}{\sqrt{\tilde\omega^2+\eta^2}}
    -\frac{1}{\sqrt{\omega^2+\eta^2}}\right)\nonumber\\
    +\int_0^{\varepsilon_c}d\omega\frac{1}{\sqrt{\omega^2+\eta^2}},
\end{eqnarray}
where the function $\tilde\omega(\omega)$ is obtained from the
continuous limit of Eq. (\ref{tilde omega eq triplet}):
\begin{equation}
\label{tilde omega eq zero T}
    \tilde\omega=\omega+\Gamma\frac{\tilde\omega}{\sqrt{\tilde\omega^2+\eta^2}}.
\end{equation}
Since $\varepsilon_c\gg\eta$, the last integral in Eq. (\ref{eta
eq triplet}) is equal to $\ln(2\varepsilon_c/\eta)$. In the clean
case we recover the BCS expression for the gap magnitude at $T=0$:
$\eta_0=(\pi/\gamma)T_{c2}=2\Gamma_c$. In the presence of disorder
Eq. (\ref{eta eq triplet}) can be represented in the form
\begin{equation}
   \ln\frac{\eta_0}{\eta}=\int_0^\infty d\omega\left(\frac{1}{\sqrt{\omega^2+\eta^2}}
    -\frac{1}{\sqrt{\tilde\omega^2+\eta^2}}\right).
\end{equation}
Using Eq. (\ref{tilde omega eq zero T}) the second term can be
transformed into an integral over $\tilde\omega$. Introducing the
notation $x=\Gamma/\eta$, we arrive at the following equation for
the gap magnitude at zero temperature as a function of $\Gamma$:
\begin{equation}
\label{eq for eta}
    {\cal F}(x)=\ln\frac{\Gamma_c}{\Gamma},
\end{equation}
where
\begin{eqnarray*}
    {\cal F}(x)=\frac{\pi x}{4}-\ln(2x)+\theta(x-1)\biggl[\ln(x+\sqrt{x^2-1})\\
    -\frac{x}{2}\arctan\sqrt{x^2-1}-\frac{\sqrt{x^2-1}}{2x}\biggr].
\end{eqnarray*}
The equation (\ref{eq for eta}) does not have solutions at
$\Gamma>\Gamma_c$, which is consistent with the complete
suppression of superconductivity above the critical disorder
strength.

Repeating the steps from the previous subsection, we obtain the
susceptibility in the form (\ref{chi ij impurities singlet}),
where
$$
    L_{ij}(\omega_n)=\Gamma
    \frac{\eta^2}{(\tilde\omega_n^2+\eta^2)^{3/2}}\langle\hat\gamma_i\hat\gamma_j\rangle_F.
$$
In the coordinate system in which
$\langle\hat\gamma_i\hat\gamma_j\rangle_F$ is diagonal, the
nonzero components of the susceptibility tensor are given by
\begin{equation}
\label{chi ii triplet}
    \frac{\chi_{ii}(T)}{\chi_P}=1-\langle\hat\gamma_i^2\rangle_F
    \pi T\sum_n\frac{\eta^2}{(\tilde\omega_n^2+\eta^2)^{3/2}
    -\Gamma\langle\hat\gamma_i^2\rangle_F\eta^2}.
\end{equation}
We note that for a spherical 3D model with
$\langle\hat\gamma_i^2\rangle_F=1/3$ this expression has exactly
the same form as the susceptibility of the superfluid ${}^3$He-B
in aerogel, see Refs. \onlinecite{MK01} and \onlinecite{SS01}.

\begin{figure}
    \includegraphics[width=\columnwidth]{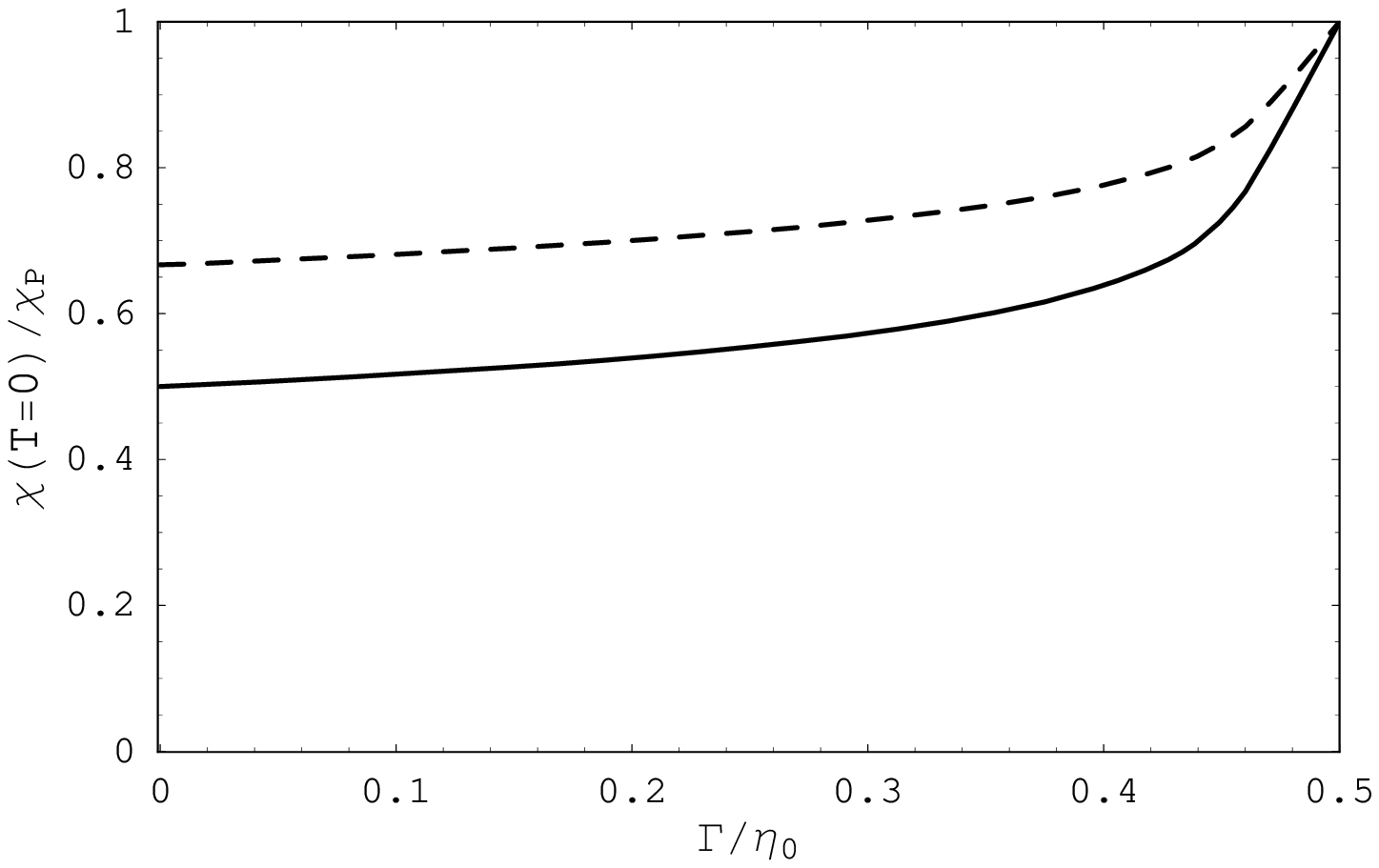}
    \caption{The residual susceptibility vs disorder strength for $\eta_+=-\eta_-$
    ($\Gamma/\eta_0=0.5$ corresponds to the critical disorder strength $\Gamma=\Gamma_c$).
    The solid line corresponds to the transverse components in the
    2D case ($\chi_{zz}=\chi_P$ and is disorder-independent), the dashed line
    -- to all three diagonal components in the 3D case.}
    \label{fig: chi_t}
\end{figure}

At zero temperature we replace the Matsubara sum by a frequency
integral and then transform it into an integral over
$\tilde\omega$ using Eq. (\ref{tilde omega eq zero T}), with the
following result:
\begin{equation}
\label{chi zero T triplet}
    \frac{\chi_{ii}(T=0)}{\chi_P}=1-\langle\hat\gamma_i^2\rangle_F
    +\langle\hat\gamma_i^2\rangle_F\Phi_2\left(\frac{\Gamma}{\eta}\right),
\end{equation}
where
\begin{eqnarray}
\label{Theta2}
    \Phi_2(x)=1-\int_{y_{min}}^\infty
    dy\left[1-\frac{x}{(y^2+1)^{3/2}}\right]\nonumber\\
    \times\frac{1}{(y^2+1)^{3/2}-x\langle\hat\gamma_i^2\rangle_F},
\end{eqnarray}
and $y_{min}=\theta(x-1)\sqrt{x^2-1}$. The last term on the
right-hand side of Eq. (\ref{chi zero T triplet}) describes the
effect of impurities on the residual susceptibility. For each
disorder strength, one first finds the solution $x(\Gamma)$ of Eq.
(\ref{eq for eta}) and then calculates $\Phi_2(x(\Gamma))$. In the
weak disorder limit we have $x(\Gamma)\simeq\Gamma/2\Gamma_c\ll
1$, and $\Phi_2(x)\simeq(3\pi
x/16)(1-\langle\hat\gamma_i^2\rangle_F)$, i.e. the residual
susceptibility increases linearly with disorder. At
$\Gamma\to\Gamma_c=\eta_0/2$:
$x(\Gamma)\simeq\sqrt{\Gamma_c/12(\Gamma_c-\Gamma)}\gg 1$. In this
limit $\Phi_2(x)\to 1$ and $\chi_{ii}(T=0)\to\chi_P$.

The dependence of $\chi_{ii}(T=0)$ on the disorder strength for
the 2D and 3D models is plotted in Fig. \ref{fig: chi_t}. As in
the case $\eta_+=\eta_-$, the residual susceptibility is enhanced
by impurities.

\section{Conclusions}
\label{sec: Conclusion}

The spin susceptibility of a clean noncentrosymmetric
superconductor has a large residual value at $T=0$, which can be
attributed to the temperature-independent contribution of the
virtual transitions between the nondegenerate bands split by the
SO coupling. We have studied the effects of disorder and found
that scalar impurities in noncentrosymmetric superconductors act
like spin-orbit impurities in centrosymmetric BCS superconductors,
in the sense that they considerably enhance the residual
susceptibility. The quantitative details depend on many factors,
in particular the shape of the Fermi surface, the structure of the
SO coupling, characterized by $\bgam(\bk)$, and the symmetry of
the order parameter.

In the model in which the gap functions in both SO split bands are
the same and isotropic, the critical temperature $T_c$ does not
depend on the elastic scattering rate $\Gamma$, while the residual
susceptibility increases and approaches the normal-state value
$\chi_P$ in the dirty limit $\Gamma\to\infty$. If the two gap
functions have the same magnitudes but opposite signs, $T_c$ is
completely suppressed above the critical concentration of
impurities corresponding to $\Gamma_c$, with the residual
susceptibility approaching $\chi_P$ at $\Gamma\to\Gamma_c$.

The band structure models considered here cannot be directly
applied to CePt$_3$Si, whose Fermi surface is quite complicated
and consists of multiple sheets.\cite{SZB04} It is not known which
one (or ones) of them are superconducting. The order parameter
symmetry is not known either, although there is strong
experimental evidence that the superconducting order parameter has
lines of gap nodes\cite{Yasuda04,Yogi04,Izawa05,Bonalde05,Tate04}
and likely corresponds to one of the nontrivial one-dimensional
representations of the point group $\mathbf{C}_{4v}$. We expect
however that the qualitative picture described above remains valid
for anisotropic pairing as well, namely the residual
susceptibility increases in the presence of impurities.

\section*{Acknowledgments}

The author is pleased to thank V. P. Mineev for stimulating
discussions. The financial support from the Natural Sciences and
Engineering Research Council of Canada and the Brock University
Research Excellence Chair Program is gratefully acknowledged.

\appendix

\section{Spin-orbit coupling of band electrons}
\label{app: SO coupling}

In order to derive the expression (\ref{H SO eff}), we start with
the Hamiltonian for non-interacting electrons in a perfect crystal
lattice:
\begin{equation}
\label{H 0}
    H_0=-\frac{\hbar^2}{2m}\bm{\nabla}^2+U(\br)
    -i\frac{\hbar^2}{4m^2c^2}\hat{\bm{\sigma}}[\bm{\nabla}U(\br)\times\bm{\nabla}],
\end{equation}
where $U(\br)$ is the lattice potential. The last term, $H_{SO}$,
represents the spin-orbit coupling. In the absence of the latter,
the eigenstates of $H_0$ are the Bloch spinors:
\begin{equation}
\label{Bloch spinors}
    \langle\br\sigma|\bk\mu\alpha\rangle=\frac{1}{\sqrt{\cal
    V}}\varphi_{\bk\mu}(\br)e^{i\bk\br}\chi_\alpha(\sigma),
\end{equation}
where ${\cal V}$ is the system volume, $\sigma$ is the spin
projection, $\varphi_{\bk\mu}(\br)$ have the same periodicity as
the crystal lattice, and $\chi_\alpha$ are the basis spinors:
$\chi_\alpha(\sigma)=\delta_{\alpha\sigma}$. The Bloch spinors are
labelled by the wave vector $\bk$, the band index $\mu$, and the
spin index $\alpha$, which distinguishes the states within the
same band. The eigenvalues $\epsilon_\mu(\bk)$ describe the
electron dispersion in the $\mu$th band and have the following
symmetry properties: $\epsilon_\mu(\bk)=\epsilon_\mu(-\bk)$,
$\epsilon_\mu(\bk)=\epsilon_\mu(g^{-1}\bk)$, where $g$ is any
operation from the point group of the crystal.

Next, we calculate the matrix elements of $H_{SO}$ in the basis of
the Bloch states (\ref{Bloch spinors}):
\begin{eqnarray}
\label{H SO matrix}
    \langle\bk\mu\alpha|H_{SO}|\bp\nu\beta\rangle=
    \frac{\hbar^2}{4m^2c^2}\sum_{ijk}e_{ijk}\langle\alpha|\hat\sigma_i|\beta\rangle\nonumber\\
    \times\frac{1}{\cal V}\int
    d^3\br\Theta_{jk}(\br)e^{i(\bp-\bk)\br},
\end{eqnarray}
where $i,j,k=x,y,z$, and
$$
    \Theta_{jk}=(\nabla_jU)\varphi^*_{\bk\mu}(-i\nabla_k+p_k)\varphi_{\bp\nu}.
$$
Since $\Theta_{jk}$ are lattice-periodic functions of $\br$, the
integral in Eq. (\ref{H SO matrix}) is nonzero only if
$\bp-\bk=\bm{G}$, where $\bm{G}$ is a reciprocal lattice vector.
Because both $\bk$ and $\bp$ are in the first Brillouin zone, the
only possibility is $\bp=\bk$. The Hamiltonian remains a
nondiagonal matrix in both the band and spin spaces, and has the
following form in the second-quantization representation:
\begin{equation}
\label{H0 gen}
    H_0=\sum_{\bk,\mu\nu,\alpha\beta}[\epsilon_\mu(\bk)\delta_{\mu\nu}\delta_{\alpha\beta}+
    \bgam_{\mu\nu}(\bk)\bm{\sigma}_{\alpha\beta}]
    a^\dagger_{\bk\mu\alpha}a_{\bk\nu\beta},
\end{equation}
where the chemical potential is included in the band dispersion
functions, and
\begin{eqnarray}
\label{gamma gen}
    &&\bgam_{\mu\nu}(\bk)=\frac{\hbar^2}{4m^2c^2}\nonumber\\
    &&\qquad\times\frac{1}{\upsilon}\int_\upsilon
    d^3\br\bigl[(\bm{\nabla}U)\times\varphi^*_{\bk\mu}
    (-i\bm{\nabla}+\bk)\varphi_{\bk\nu}\bigr]\qquad
\end{eqnarray}
(the integration is performed over the unit cell of volume
$\upsilon$). The expression (\ref{H0 gen}) is exact for
non-interacting electrons, regardless of the band structure and
the strength of the SO coupling.

The functions $\bgam_{\mu\nu}(\bk)$ satisfy certain
symmetry-imposed conditions. Since $H_0$ is Hermitian, we have
\begin{equation}
\label{gamma herm}
    \bgam_{\mu\nu}(\bk)=\bgam^*_{\nu\mu}(\bk).
\end{equation}
Under the point group operations $g$, e.g. rotations, the
second-quantization operators transform as follows:
$a^\dagger_{\bk\mu\alpha}\to\sum_\beta
a^\dagger_{g\bk,\mu\beta}{\cal U}_{\beta\alpha}(g)$, where
$\hat{\cal U}(g)=e^{-i\theta(\bm{n}\hat{\bm{\sigma}})/2}$ is the
spinor representation of the rotation about a direction $\bm{n}$
by an angle $\theta$.\cite{ED-Book} Requiring that $H_0$ remains
invariant under $g$ we obtain:
\begin{equation}
\label{gamma R}
    \bgam_{\mu\nu}(\bk)=g\bgam_{\mu\nu}(g^{-1}\bk).
\end{equation}
Under time reversal, $fa^\dagger_{\bk\mu\alpha}\to
f^*\sum_\beta(i\sigma_2)_{\alpha\beta}a^\dagger_{-\bk\mu\beta}$
($f$ is an arbitrary $c$-number coefficient), therefore
$\bgam_{\mu\nu}(\bk)\to-\bgam_{\mu\nu}^*(-\bk)$. If the
time-reversal symmetry is not broken, we have
\begin{equation}
\label{gamma K}
    \bgam_{\mu\nu}(\bk)=-\bgam_{\mu\nu}^*(-\bk).
\end{equation}
Finally, under inversion $a^\dagger_{\bk\mu\alpha}\to
a^\dagger_{-\bk\mu\alpha}$, and $\bgam_{\mu\nu}$ transform like
pseudovectors: $\bgam_{\mu\nu}(\bk)\to\bgam_{\mu\nu}(-\bk)$.

In a centrosymmetric crystal,
$\bgam_{\mu\nu}(\bk)=\bgam_{\mu\nu}(-\bk)$, therefore, using Eq.
(\ref{gamma K}),
\begin{equation}
\label{gamma I}
    \bgam_{\mu\nu}(\bk)=-\bgam_{\mu\nu}^*(\bk).
\end{equation}
It follows from the conditions (\ref{gamma herm}) and (\ref{gamma
I}) that $\bgam_{\mu\mu}(\bk)=0$. Therefore one needs to include
at least two bands in Eq. (\ref{H0 gen}), in which case
$\bgam_{12}(\bk)=-\bgam_{21}(\bk)=i\bm{\ell}(\bk)$, and the
Hamiltonian takes the form
\begin{eqnarray}
\label{H0-cs}
    H_0&=&\sum_{\bk,\alpha}\sum_{\mu=1,2}\epsilon_\mu(\bk)a^\dagger_{\bk\mu\alpha}
    a_{\bk\mu\alpha}\nonumber\\
    &&+i\sum_{\bk,\alpha\beta}\bm{\ell}(\bk)\bm{\sigma}_{\alpha\beta}
    (a^\dagger_{\bk 1\alpha}a_{\bk 2\beta}-a^\dagger_{\bk 2\alpha}a_{\bk
    1\beta}),\qquad
\end{eqnarray}
see also Ref. \onlinecite{Gor65}. The pseudovector $\bm{\ell}$ is
real, even in $\bk$, and satisfies
$\bm{\ell}(\bk)=g\bm{\ell}(g^{-1}\bk)$.

In contrast, in a noncentrosymmetric crystal, the constraint
(\ref{gamma I}) is absent, and the effects of SO coupling can be
studied in a minimal model in which one keeps just one band in the
Hamiltonian (\ref{H0 gen}). The band index $\mu$ can then be
dropped and the SO coupling can be described by a single
pseudovector function $\bgam(\bk)$, which is real, odd in $\bk$,
and invariant with respect to the point group operations:
$\bgam(\bk)=g\bgam(g^{-1}\bk)$. In this way one arrives at the
effective band Hamiltonian (\ref{H SO eff}). Although in principle
one can calculate $\bgam(\bk)$ using Eq. (\ref{gamma gen}), we
consider it as a model parameter.


\begin{thebibliography}{99}
\bibitem{negative signs}
The origin of the negative signs in Eqs. (\ref{spin-singlet}) and
(\ref{spin-triplet}) can be traced back to our definition of the
time reversal operator: $K=i\hat\sigma_2K_0$ ($K_0$ is the complex
conjugation). If one uses $K=-i\hat\sigma_2K_0$ instead, then
$\psi=(\tilde\Delta_++\tilde\Delta_-)/2$ and
$\bm{d}=(\tilde\Delta_+-\tilde\Delta_-)\hat{\bgam}/2$, as in Ref.
\onlinecite{Min04}.
\end{thebibliography}
\end{document}